\documentclass[useAMS,usenatbib]{mnras}
\usepackage{graphicx}
\usepackage{amsmath, amssymb}
\usepackage[T1]{fontenc}
\usepackage[usenames]{color}
\usepackage[dvipsnames]{xcolor}
\usepackage{bm}
\usepackage[capitalize]{cleveref}
\usepackage{physics}
\usepackage{bm}
\usepackage{acro}
\usepackage{caption}
\usepackage{subcaption}
\usepackage[normalem]{ulem}

\DeclareAcronym{DM}{short = DM, long  = dark matter}
\DeclareAcronym{IC}{short = IC, long  = initial condition}
\DeclareAcronym{NGP}{short = NGP, long  = nearest grid point}
\DeclareAcronym{MW}{short = MW, long  = Milky Way}
\DeclareAcronym{BORG}{short = \texttt{BORG}, long  = \textit{Bayesian Origin Reconstruction from Galaxies}}
\DeclareAcronym{CSIBORG}{short = \texttt{CSiBORG}, long  = \textit{Constrained Simulations in} \texttt{BORG}}
\DeclareAcronym{FOF}{short = \texttt{FOF}, long  = friends-of-friends}
\DeclareAcronym{WMAP}{short = WMAP, long  = Wilkinson Microwave Anisotropy Probe}
\DeclareAcronym{HMF}{short = HMF, long  = halo mass function}
\DeclareAcronym{FWHM}{short = FWHM, long  = full width at half maximum}
\DeclareAcronym{SHAM}{short = SHAM, long  = subhalo abundance matching}

\title[Variance of haloes in constrained simulations]{Evaluating the variance of individual halo properties in constrained cosmological simulations}

\author[R. Stiskalek et al.]{
Richard Stiskalek,$^{1}$\thanks{\href{mailto:richard.stiskalek@physics.ox.ac.uk.com}{richard.stiskalek@physics.ox.ac.uk}}
Harry Desmond,$^{2}$
Julien Devriendt$^{1}$
and Adrianne Slyz$^{1}$
\\
$^{1}${Department of Physics, University of Oxford, Denys Wilkinson Building, Keble Road, Oxford OX1 3RH, United Kingdom}\\
$^{2}$Institute of Cosmology \& Gravitation, University of Portsmouth, Dennis Sciama Building, Portsmouth, PO1 3FX, UK
}

\date{Accepted XXX. Received YYY; in original form ZZZ}
\pubyear{2023}

\begin{document}\label{firstpage}
\pagerange{\pageref{firstpage}--\pageref{lastpage}}
\maketitle

\begin{abstract}
Constrained cosmological simulations play an important role in modelling the local Universe, enabling investigation of the dark matter content of local structures and their formation. We introduce an internal method for quantifying the extent to which the variance of individual halo properties is suppressed by the constraints imposed on the initial conditions. We apply it to the \textit{Constrained Simulations in BORG} (\texttt{CSiBORG}) suite of $101$ high-resolution realisations across the posterior probability distribution of initial conditions from the \textit{Bayesian Origin Reconstruction from Galaxies} (\texttt{BORG}) algorithm. The method is based on the overlap of the initial Lagrangian patch of a halo in one simulation with those in another, measuring the degree to which the haloes' particles are initially coincident. This addresses the extent to which the imposed large-scale structure constraints reduce the variance of individual halo properties. We find consistent reconstructions of $M\gtrsim10^{14}~M_\odot / h$ haloes, indicating that the constraints from the \texttt{BORG} algorithm are sufficient to pin down the masses, positions, and peculiar velocities of clusters to high precision, though we do not assess how well they reproduce observations of the local Universe. The effect of the constraints tapers off towards lower mass, and the halo spins and concentrations are largely unconstrained at all masses. We document the advantages of evaluating halo consistency in the initial conditions and describe how the method may be used to quantify our knowledge of the halo field given galaxy survey data analysed through the lens of probabilistic inference machines such as \texttt{BORG}.
\end{abstract}



\section{Introduction}\label{sec:intro}

The dynamics of the Universe are largely governed by \ac{DM}, constituting the majority of the matter content of the Universe. Over the past decades, cosmological simulations have emerged as the paramount instrument to elucidate its nonlinear dynamics and interplay with baryons~\citep{Wechsler_2018, Vogelsberger_2020, Angulo_2022}. Simulations typically employ \acp{IC} based on a realisation of a $\Lambda\mathrm{CDM}$ power spectrum and random phases of the primordial matter field~\citep{Press_1974,Davis1985_FoF, Lacey_1993,Eisenstein1998,Tinker_2008}. Such \acp{IC} produce universes that resemble the real Universe only statistically, but cannot be linked object-by-object.

The alternative is the ``\emph{constrained simulation}'', in which not only the amplitudes but also the phases of the primordial density perturbations are encoded. The beginnings of this endeavour can be traced back to~\cite{Bertschinger_1987} and~\cite{Hoffman_1991}, who laid the foundation for simulating constrained realisations of Gaussian random fields. Local Universe constraints were subsequently derived from galaxy counts~\citep{Kolatt_1996, Bistolas_1998}, peculiar velocity measurements~\citep{Weygaert_2000, Klypin_2003, Kravtsov_2002} and galaxy groups~\citep{Wang_2014}, which, together with advances in simulation resolution, gravity modelling, \ac{IC} generation or galaxy bias modelling, have led to local Universe simulation becoming a mature field.

Deriving the \acp{IC} that generated the structure we see around us is an inference problem, and, as we have only one Universe, is best formulated in a Bayesian framework. This realisation led to the development of a Bayesian forward-modelling approach now known as the \ac{BORG} algorithm~\citep{Jasche_2013,Jasche_2015,Lavaux_2016_2MPP,Jasche2019_BORG}, although other early studies explored the forward-modelling approach as well (e.g.~\citealt{Kitaura_2013, Hess_2013, Wang_2013}).~\ac{BORG} leverages an efficient Hamiltonian Markov Chain Monte Carlo algorithm to sample the initial matter field along with parameters associated with observational selection effects, galaxy bias, and cosmology. The \ac{BORG} posterior encapsulates all realisations of the local Universe that are compatible with the observational constraints used to derive them. The flagship application of \ac{BORG}---and the one that we will use in this work---targeted the $\mathrm{2M}\texttt{++}$ galaxy catalogue, a whole-sky redshift compilation of $69,160$ galaxies~\citep{Lavaux_2016_2MPP,Jasche2019_BORG} based on the Two-Micron-All-Sky Extended Source Catalog~\citep{2MASS}. Work is in progress to augment the constraints with information from cosmic shear \citep{Porqueres_2021} and peculiar velocities \citep{Prideaux-Ghee_2023}.

In this work, we use the \ac{CSIBORG} suite of constrained cosmological simulations~\citep{Bartlett_2021,Desmond2022_antihalo}, which are based on the \ac{BORG} $\mathrm{2M}\texttt{++}$ \acp{IC}. Each \ac{CSIBORG} box is a resimulation of \acp{IC} from a single \ac{BORG} posterior sample inferred from the $\mathrm{2M}\texttt{++}$ galaxy catalogue~\citep{Lavaux_2016_2MPP,Jasche2019_BORG}, so that differences between realisations quantify the reconstruction uncertainty associated with our incomplete knowledge of the galaxy field and galaxy--halo connection. \cite{Max2022} used \ac{CSIBORG} to study the effect of the reduced cosmic variance on the \ac{HMF} and clustering of haloes, and developed a method to assess the consistency of halo reconstruction from the final conditions. \ac{CSIBORG} has also previously been used to create catalogues of local voids-as-antihaloes~\citep{Desmond2022_antihalo}, and search for modified gravity~\citep{Bartlett_2021} and dark matter annihilation and decay~\citep{Bartlett_2022, Kostic_2023}.

A focus of constrained simulations has been used to study the Local Group and its assembly history, e.g., within the \texttt{CLUES}~\citep{Gottloeber_2010,Sorce_2015} and \texttt{SIBELIUS}~\citep{Sawala_2022,McAlpine_2022_Sibellius} projects. The \texttt{CLUES} collaboration explored the reconstruction of the local Universe and its clusters (e.g.~\citealt{Sorce_2016, Sorce_2016_Virgo, Sorce_2020B}). Constrained simulations have also been used to study the connection between Sloan Digital Sky Survey galaxies and their haloes~\citep{Wang_2016, Yang_2018, Zhang_2022, Xu_2023}, quantify the compatibility of the local Universe with $\Lambda\mathrm{CDM}$~\citep{Stopyra_2021, Stopyra_2023}, model the local Universe in modified gravity~\citep{Naidoo_2023}, model the reionization of the local Universe (e.g.,~\citealt{Ocvirk_2016, Aubert_2018, Ocvirk_2020}), or predict the future evolution of protoclusters~\citep{Ata_2022}. Recently, the \texttt{SLOW} project was introduced, which aims to resimulate constrained \acp{IC} from the Constrained LOcal and Nesting Environment Simulations project (\texttt{CLONES};~\citealt{Sorce_2018B}) using a hydrodynamical simulation~\citep{Dolag_2023, Boss_2023, Martinez_2024}.

Due to their Bayesian setup, \ac{BORG} and \ac{CSIBORG} afford quantification of the effects of data and model uncertainties on the dark matter distribution produced in the simulations. An alternative method leverages the Wiener filter, which allows reconstruction of the mean field but cannot quantify reconstruction uncertainty~\citep{Hoffman_1991, Zaroubi_1995,Zaroubi_1999, Doumler_2013A,Doumler_2013B, Hoffman_2015}. In a recent study,~\cite{Valade_2023} compared a Wiener filter and Bayesian-based (\texttt{HAMLET}; \citealt{Valade_2022_HAMLET}) approach to reconstruction from a mock peculiar velocity catalogue. For example, to quantify reconstruction uncertainty, the \texttt{CLUES} team (as described in \citealt{Sorce_2014, Sorce_2016}) uses constrained realizations to estimate the possible residuals between the Wiener filter and the true underlying field~\citep{Hoffman_1991, Hoffman_1992}. In this approach, they add random power at large scales---if the Wiener filter is not constrained---to ensure that the resulting power spectrum matches the underlying cosmological model. By varying the random seed for this added power, they can quantify the impact of unconstrained modes on the reconstruction~\citep{Sorce_2016_Virgo, Sorce_2020}. Closely related to this work,~\cite{Sorce_2019} investigated the suppression of cosmic variance in constrained simulations of the Virgo cluster, comparing the reconstructed properties to observational estimates and finding good agreement. Additionally,~\citeauthor{Sorce_2019} quantified the uniqueness of the Virgo cluster in comparison to a population of random haloes.

The objective of our study is to investigate precision or robustness of the reconstructions of individual haloes in constrained cosmological simulations, as opposed to accuracy in the sense of how well the reconstructions match the local Universe. We develop a framework to assess whether a halo present in one box is also present in another, and hence what properties of the haloes are robustly reconstructed across the suite, in the sense of being insensitive to simulation within the suite. This is achieved by means of a novel metric, the overlap of haloes' initial Lagrangian patches. While the method is agnostic as to the way in which the simulations of the suite are linked, a natural application (and the one on which we focus) is to suites that sample the \ac{IC} posterior of a previous inference (e.g.,~\ac{BORG}). We showcase the method by application to \ac{CSIBORG}, where we quantify the consistency of the halo reconstruction as a function of various halo properties. The significance of the results are established by contrast with \texttt{Quijote}, an unconstrained suite. This metric quantifies the consistency of halo reconstruction (precision), but it does not assess how reliably these halos match onto structures in the real local Universe (accuracy).

Other applications of our method include matching haloes between \ac{DM}-only simulations and their hydrodynamical counterparts, as well as simulations using different cosmological models, e.g. $\Lambda\mathrm{CDM}$ and modified gravity. While traditional methods for matching haloes between different runs often depend on a consistent \ac{DM} particle ordering between the runs (e.g.~\citealt{Butsky_2016, Desmond_2017, Mitchell_2018, Cataldi_2021}), ours does not. In both above-mentioned scenarios, our approach can quantify how either the hydrodynamics or the cosmology impacts the properties of individual haloes, instead of relying solely on population statistics conditioned on properties such as mass (e.g.~\citealt{Pallero_2023}).

The structure of the paper is as follows. In~\Cref{sec:simdata} we introduce the two sets of simulations employed in our work, in~\Cref{sec:method} we introduce the overlap metric and its interpretation, ~\Cref{sec:results} contains our results and in~\Cref{sec:discussion} we discuss the results. Lastly, we conclude in~\Cref{sec:conclusion}. All logarithms in this work are base-10.


\section{Simulated data}\label{sec:simdata}

In this section, we describe the two sets of simulations that we use, \ac{CSIBORG} and \texttt{Quijote}, and their halo catalogues.

\begin{figure}
    \centering
    \includegraphics[width=\columnwidth]{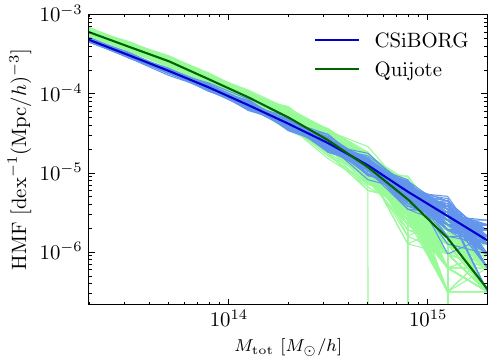}
    \caption{The \ac{FOF} \ac{HMF} in \ac{CSIBORG} and \texttt{Quijote}. The thin background lines show the realisations of \ac{CSIBORG} and \texttt{Quijote} and the bold lines show their mean. The \ac{CSIBORG} \ac{HMF} undershoots and overshoots the \texttt{Quijote} \ac{HMF} at low and high masses, respectively. The systematic offset is driven by the \ac{BORG} gravity model and is not associated with deviations of the local Universe from the cosmic average (see~\cref{sec:simdata_halocat} for further discussion). The lower limit is approximately given by \texttt{Quijote} haloes containing $100$ particles.}
    \label{fig:HMF_comparison}
\end{figure}


\subsection{CSiBORG}\label{sec:simdata_csiborg}

The \ac{CSIBORG} suite, first presented in~\citet{Bartlett_2021}, consists of $101$ \ac{DM}-only $N$-body simulations in a $677.7~\mathrm{Mpc}/h$ box centred on the \ac{MW}, with \acp{IC} sampled from the \ac{BORG} reconstruction of the $\mathrm{2M}\texttt{++}$ galaxy survey~\citep{Lavaux_2016_2MPP}. This reconstruction covers the same volume as each \ac{CSIBORG} box, discretised into $256^3$ cells for a spatial resolution of $2.65~\mathrm{Mpc}/h$~\citep{Jasche2019_BORG}. The \ac{BORG} density field is constrained within a spherical volume of radius $\sim155~\mathrm{Mpc}/h$ around the \ac{MW}, where the $\mathrm{2M}\texttt{++}$ catalogue has high completeness.

In \ac{CSIBORG}, the \acp{IC} are propagated linearly to $z = 69$ and augmented with white noise on a $2048^3$ grid in the central high-completeness region, corresponding to a spatial resolution of $0.33~\mathrm{Mpc}/h$ and a \ac{DM} particle mass of $3.09 \times 10^9~\mathrm{M}_\odot/h$. To ensure a smooth transition to the remainder of the box, a buffer region of approximately $10~\mathrm{Mpc}/h$ is added at the edge of the high-resolution region. Both \ac{BORG} and \ac{CSIBORG} adopt the cosmological parameters from the~\cite{Planck2014} best fit results, including the \ac{WMAP} polarisation, high multipole moment, and baryonic acoustic oscillation data, except $H_0$ which is taken from the 5-year \ac{WMAP} results combined with Type Ia supernovae and baryonic acoustic oscillation data~\citep{Hinshaw2009} ($T_\mathrm{CMB} = 2.728~\mathrm{K}$, $\Omega_\mathrm{m} = 0.307$, $\Omega_\Lambda = 0.693$, $\Omega_\mathrm{b} = 0.04825$, $H_0 = 70.5~\mathrm{km}~\mathrm{s}^{-1}~\mathrm{Mpc}^{-1}$, $\sigma_8 = 0.8288$, $n = 0.9611$). The \ac{DM} density field is evolved to $z = 0$ using the adaptive mesh refinement code RAMSES~\citep{Teyssier2002_RAMSES}, where only the central high-resolution region is refined (reaching level $18$ by $z = 0$ with a spatial resolution of $2.6~\mathrm{kpc} / h$).

As we will be studying the reconstruction of individual objects, whose initial Lagrangian patches are constrained in \ac{BORG}, it is illustrative to consider how many \ac{BORG} cells constitute the Lagrangian patch of a halo. We find this to be approximately $N \approx 7~M_{\rm tot} / (10^{13} M_\odot / h)$. \ac{BORG} constrains the average field value in each cell with physical size $2.65~\mathrm{Mpc} / h$, and we do not a priori expect haloes with Lagrangian patches spanning only a few \ac{BORG} cells to be consistently reconstructed in \ac{CSIBORG} since such haloes likely vary strongly across the \ac{BORG} posterior. However, haloes above $\sim 10^{14}~\mathrm{Mpc} / h$ comprise initially $\gtrsim 100$ cells and are therefore likely well constrained.


\subsection{Quijote}\label{sec:simdata_quijote}

We compare the \ac{CSIBORG} results to the publicly available \texttt{Quijote} simulations\footnote{\url{https://quijote-simulations.readthedocs.io/}}~\citep{Quijote_sims}. \texttt{Quijote} is a suite of unconstrained simulations evolved from $z = 127$ to $z = 0$ using the \texttt{GADGET-III} code~\citep{Springel2008_GADGET3}. We use $10$ realisations of the \texttt{Quijote} \ac{DM}-only simulations with randomly drawn \ac{IC} phases, each with a volume of $1~\mathrm{Gpc}/h$ and a particle mass of $8.72 \times 10^{10}~\mathrm{M}_\odot/h$ in a fiducial cosmology: $\Omega_\mathrm{m} = 0.3175$, $\Omega_\Lambda = 0.6825$, $\Omega_\mathrm{b} = 0.049$, $H_0 = 67.11~\mathrm{km}~\mathrm{s}^{-1}~\mathrm{Mpc}^{-1}$, $\sigma_8 = 0.834$ and $n = 0.9624$. Besides the constraints, the most significant difference with \ac{CSIBORG} is the volume, so for an approximately fair comparison when calculating any extrinsic quantity we mimic the \ac{CSIBORG} high-resolution region by splitting each \texttt{Quijote} box into $27$ non-overlapping spherical sub-volumes of radius $155~\mathrm{Mpc} / h$ centred at $n \times 155~\mathrm{Mpc} / h$ for $n=1, 3, 5$ along each axis.

\subsection{Halo catalogues}\label{sec:simdata_halocat}

We use the \acl{FOF} halo finder (\acs{FOF};~\citealt{Davis1985_FoF}) in both \ac{CSIBORG} and \texttt{Quijote}, with a linking length parameter of $b = 0.2$. \ac{FOF} connects particles within a distance $b$ times the mean interparticle separation. FOF can create artificially large structures by connecting extraneous particles to haloes along nearby filaments of the density field, particularly for merging haloes at $z = 0$~\citep{Eisenstein1998}. \cite{Warren2006} and~\cite{Lukic2009} proposed corrections to this based on particle number and halo concentration, respectively, but as we do not require high-precision halo masses we do not apply them here.

To reduce numerical resolution errors of recovered haloes and their properties, the authors have suggested that haloes must contain at least $50-100$ particles~\citep{Springel2008_GADGET3,Onions2012_subhalos,Knebe2013,Bosch2016,Griffen2016}, though~\cite{Diemer_2015} suggested stricter criteria for measuring e.g. concentration to avoid having only a few particles in the inner parts of the halo, especially if the inner region spans only a few force resolution lengths. Recently,~\cite{Bosch2018} proposed a more stringent criterion to determine the numerical convergence of infalling subhaloes. However, in this work, we are not concerned with substructure or halo profiles and therefore adopt the simpler and less restrictive criterion of $100$ particles for all haloes, corresponding to a minimum halo mass of $3.09 \times 10^{11}~\mathrm{M}_\odot/h$ and $8.72 \times 10^{12}~\mathrm{M}_\odot/h$ for \ac{CSIBORG} and \texttt{Quijote}, respectively. In \ac{CSIBORG} we identify haloes only inside the high-resolution region. As \ac{CSIBORG} has a mass resolution nearly two orders of magnitude better than \texttt{Quijote}, when comparing the two simulations, we only consider haloes above the mass resolution of \texttt{Quijote}.

\ac{CSIBORG} and \texttt{Quijote} assume different cosmological parameters, yielding a difference in both the large-scale structure and halo population.~\cite{Max2022} study how this affects the \ac{HMF}, and show in their Fig. 1 no significant disparities due to the different cosmologies. They do however quantify the reduced spread of the \acp{HMF} of individual \ac{CSIBORG} realisations relative to \texttt{Quijote} due to the suppression of cosmic variance by the \ac{IC} constraints. We show the \acp{HMF} in \cref{fig:HMF_comparison}. It is found that the \ac{CSIBORG} and \texttt{Quijote} HMFs have some systematic disagreement, with the former predicting higher bright-end abundances (see also \citealt{McAlpine_2022_Sibellius,Desmond2022_antihalo,Max2022}). However, this becomes statistically significant only above $M \sim 10^{15} M_\odot / h$. On the other hand, \ac{CSIBORG} systematically undershoots the \texttt{Quijote} \ac{HMF} below $\sim 10^{14.2}~M_\odot / h$. These discrepancies have recently been investigated by~\cite{Stopyra_2023} who showed that replacing the $10$-step particle mesh solver used in the \ac{BORG} forward model with a $20$-step COLA solver~\citep{Tassev_2013_COLA} corrects for them.


\section{Methodology}\label{sec:method}

\begin{figure*}
    \centering
    \includegraphics[width=\textwidth]{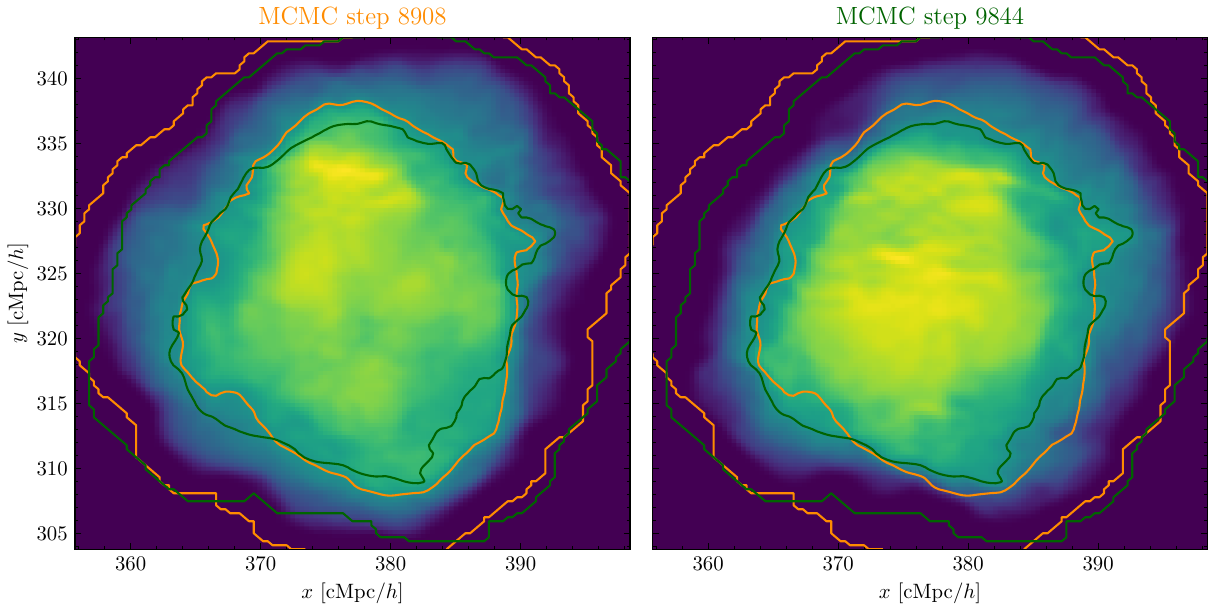}
    \caption{Example of Lagrangian patches at $z = 69$ for two haloes identified at $z = 0$ from \ac{CSIBORG} at two distinct Markov Chain Monte Carlo steps of the \ac{BORG} posterior. Both haloes have a mass of $10^{15} ~ M_\odot / h$ at $z = 0$ and a mutual Lagrangian patch overlap of approximately $75\%$. The plot shows the projected mass density of the Lagrangian patches in the $x$-$y$ plane, along with density contours corresponding to the two haloes. The contour colours match that of the labels above the figures. Notably, the size of each Lagrangian patch of such haloes is around $30~\mathrm{cMpc} / h$. Both panels share the same $x$ and $y$ axes.}
    \label{fig:lagpatch_example}
\end{figure*}

We aim to develop a metric to quantify the precision with which haloes are robustly reconstructed between boxes, e.g. across the simulations of constrained suites. We do so by evaluating the similarity of proto-haloes at high redshift, measured by their initial Lagrangian patches. This is a priori a sensible approach as it is the initial conditions that are constrained: focusing on the initial snapshot facilitates the establishment of a more causally coherent framework, circumventing reliance on interpretations deduced purely from the final conditions as in e.g.~\cite{Max2022}. Most of our work is intended to interpret and show that it is useful a posteriori as well.

Similar approaches are used to associate haloes between \ac{DM}-only and hydrodynamical simulations sharing the same \acp{IC} (e.g.~\citealt{Butsky_2016, Desmond_2017, Mitchell_2018,Cataldi_2021}). For instance,~\citeauthor{Desmond_2017} leverage the ability to match \ac{DM} particle IDs between \ac{DM}-only and hydrodynamical runs to match haloes as follows: If halo $a$ from the \ac{DM}-only run shares the most particles with halo $b$ from the hydrodynamical run, and conversely, $b$ shares most particles with $a$, they are identified as a match. This bears a strong resemblance to the method we develop because the particle IDs are assigned based on their position in the initial snapshot. In \ac{CSIBORG} the particle IDs are not consistent between realisations, so this method cannot be used directly.

We cross-correlate two \ac{IC} realisations as follows. We denote the first simulation as the ``reference'' and the second as the ``crossing'' simulation and calculate the overlap of each halo in the reference simulation with all haloes in the crossing simulation. We identify \ac{FOF} haloes in the final snapshot of both the reference $\mathcal{A}$\textsuperscript{th} and the crossing $\mathcal{B}$\textsuperscript{th} \ac{IC} realisation and trace their constituent particles back to the initial snapshot. To calculate the intersecting mass of a halo $a \in \mathcal{A}$ and $b \in \mathcal{B}$, we  use the \ac{NGP} scheme to assign the halo particles to a $2048^3$ grid in the initial snapshot, matching the initial refinement of the high-resolution region. We denote the mass assigned to the $n$\textsuperscript{th} cell as $m_a^{(n)}$ and $m_b^{(n)}$, respectively, for the two haloes. On the same grid, we also calculate the background mass field of all particles assigned to a halo in the final snapshot in the respective simulation, $\widehat{m}_{\mathcal{A}}^{(n)}$ and $\widehat{m}_{\mathcal{B}}^{(n)}$.

In the \ac{NGP} scheme, a particle located at $x_i \in \left[0, 1\right]$ along the $i$\textsuperscript{th} axis is assigned to the $\lfloor x_i N_{\rm cells}\rfloor$\textsuperscript{th} cell, where $N_{\rm cells}$ is the total number of cells along an axis. We then apply a Gaussian smoothing to the \ac{NGP} field using the kernel width of one cell and define the intersecting mass of haloes $a$ and $b$ as
\begin{equation}\label{eq:intersect}
    X_{a b}
    \equiv
    \sum_{n} \frac{2 m_{a}^{(n)} m_{b}^{(n)}}{\widehat{m}_{\mathcal{A}}^{(n)} + \widehat{m}_{\mathcal{B}}^{(n)}},
\end{equation}
where $n = 1,\ldots,2048^3$ is the grid index. This definition accounts for the fact that particles of more than one halo may contribute to a single cell and may be motivated as
\begin{equation}
    X_{a b} = \sum_n\left[
    \left(\frac{m_{b}^{(n)}}{\widehat{m}_{\mathcal{A}}^{(n)} + \widehat{m}_{\mathcal{B}}^{(n)}}\right) m_{a}^{(n)}
    +
    \left(\frac{m_{a}^{(n)}}{\widehat{m}_{\mathcal{A}}^{(n)} + \widehat{m}_{\mathcal{B}}^{(n)}}\right) m_{b}^{(n)}\right],
\end{equation}
i.e. the contribution of $a$ is weighted by the mass of $b$ in that cell normalized by the total mass in that cell, and vice versa. Next, we define the \ac{IC} overlap between the haloes $a$ and $b$ as
\begin{equation}\label{eq:overlap}
    \mathcal{O}_{a b} = \frac{X_{a b}}{{M_a} + M_{b} - X_{a b}},
\end{equation}
such that $M_{a} = \sum_{n} m_{a}^{(n)}$ is the total particle mass of the $a$\textsuperscript{th} halo, and similarly for the $b$\textsuperscript{th} halo. This ensures that $\mathcal{O}_{a b} \in \left[0, 1\right]$ and can be interpreted simply as the mass of the $a$\textsuperscript{th} halo that overlaps with the $b$\textsuperscript{th} halo normalised by their total mass. The definition of~\cref{eq:intersect} accounts for the fact that some cells of a halo $a \in \mathcal{A}$ may overlap simultaneously with haloes $b_1, b_2 \in \mathcal{B}$. Due to the presence of $b_2$, the intersecting mass $X_{a b_1}$ is appropriately reduced (denominator of Eq.~\ref{eq:intersect}) to avoid double-counting. For further illustration, in~\cref{fig:lagpatch_example}, we present an example of Lagrangian patches from two haloes in \ac{CSIBORG} with a $75\%$ overlap. Despite being drawn from distinct steps of the \ac{BORG} posterior, the two Lagrangian patches are located at the same comoving position and lead to a collapsed structure at $z = 0$ with a mass of approximately $10^{15} ~ M_\odot / h$, also at a similar location.

We can illustrate the overlap on a toy example of two haloes. If we have that $m_a^{(n)} = m_b^{(n)} = \widehat{m}_\mathcal{A}^{(n)} = \widehat{m}_\mathcal{B}^{(n)} = m$ in all cells, then $X_{a b}$ is simply $m$ times the number of cells occupied simultaneously by both $a$ and $b$. Moreover, if the $a$\textsuperscript{th} halo is completely enclosed within the $b$\textsuperscript{th} halo, then the overlap can be expressed as $\mathcal{O}_{a b} = M_a / M_b$, i.e. the ratio of their masses. In case of perfect overlap $M_a = M_b$ and thus $\mathcal{O}_{a b} = 1$.

The overlap measures the similarity of two haloes in their Lagrangian patches. However, a halo may overlap with many smaller haloes, producing a large set of small overlaps with none of the overlapping haloes being similar to the reference halo in mass or size. Therefore, we also calculate the maximum overlap that a halo $a$ has with any halo in the crossing simulation $\max_{b \in \mathcal{B}} \mathcal{O}_{a b}$. If this quantity is sufficiently high across all crossing simulations, it implies that the halo is consistently reconstructed across the \ac{IC} realisations. While the overlap between a pair of haloes is symmetric, if a halo $a_0 \in \mathcal{A}$ has a maximum overlap $\mathcal{O}_{a_0 b_0} = \max_{b \in \mathcal{B}} \mathcal{O}_{a_0 b}$ with some halo $b_0 \in \mathcal{B}$, this does not imply that $b_0$ also has a maximum overlap with $a_0$ since its maximum overlap is defined as $\max_{a \in \mathcal{A}} \mathcal{O}_{b_0 a}$ and the two sets of haloes over which the overlaps are maximised are not the same.

The properties of the overlap $\mathcal{O}_{a b}$ lend it a natural interpretation as the probability of a match between haloes in two simulations. That a reference halo can have a non-zero overlap with multiple haloes that themselves may overlap in their initial Lagrangian patches is already accounted for in the definition of the overlap through the denominator of the intersecting mass in~\cref{eq:intersect}, which implies that the overlap of a pair of haloes is modified by the presence of other overlapping haloes. Therefore, we consider that the probability that a reference halo $a$ being matched to \emph{some} halo in the crossing simulation is simply the sum of the overlaps with all haloes in the crossing simulation:
\begin{equation}\label{eq:prob_match}
    P(\mathrm{a} \in \mathcal{A} ~\text{matched in}~\mathcal{B})
    =
    \sum_{b \in \mathcal{B}} \mathcal{O}_{a b},
\end{equation}
and the probability of not being matched to any halo in the crossing simulation is
\begin{equation}\label{eq:prob_not_match}
    P(\mathrm{a} \in \mathcal{A} ~\text{not matched in}~\mathcal{B})
    =
    1 - \sum_{b \in \mathcal{B}} \mathcal{O}_{a b}.
\end{equation}
The definitions of the intersecting mass and overlap in~\cref{eq:intersect,eq:overlap} ensure that not only the overlap between a pair of haloes is always $\le 1$, but also that the sum of overlaps with a reference halo such as in~\cref{eq:prob_match} is always $\le 1$. In the denominator of~\cref{eq:intersect}, the intersecting mass is weighted by the fraction of a cell occupied by the pair of haloes, taking into account the contributions from all haloes in both simulations. This ensures that the intersecting mass is not counted multiple times when summing over crossing haloes.

We calculate the most likely value of a matched halo property $\mathcal{H}$ (for instance, mass) based on haloes that overlap with the reference halo. For each crossing simulation indexed $i$ we select $\mathcal{H}_i$ of the halo with the highest overlap $w_i$ with the reference halo. We then calculate the most likely matched property as the mode of the distribution of $\mathcal{H}_i$ weighted by $w_i$. To identify it, we employ the shrinking sphere method, commonly utilised for finding the centre of \ac{DM} haloes~\citep{Power_2003}. We iteratively shrink the search radius around the weighted average of the enclosed $\mathcal{H}_i$ samples until less than $5$ samples are enclosed, at which point we take their average as the mode of the distribution. We verify that $5$ samples are sufficient and our conclusions are not affected by increasing it.


\section{Results}\label{sec:results}

\begin{figure*}
    \begin{subfigure}[t]{0.48\textwidth}
        \includegraphics[width=\linewidth]{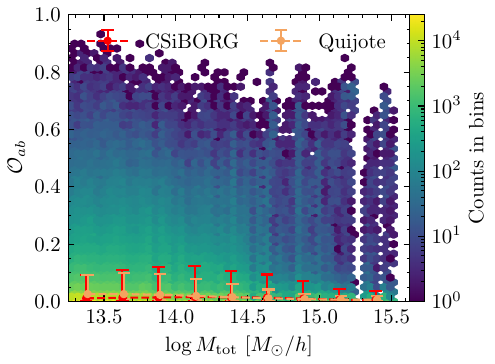}
        \caption{Concatenated pair overlaps $\mathcal{O}_{a b}$ between a reference halo and all haloes from all remaining simulations.}
        \label{fig:pair_overlap}
    \end{subfigure}
    \hfill
    \begin{subfigure}[t]{0.48\textwidth}
        \includegraphics[width=\linewidth]{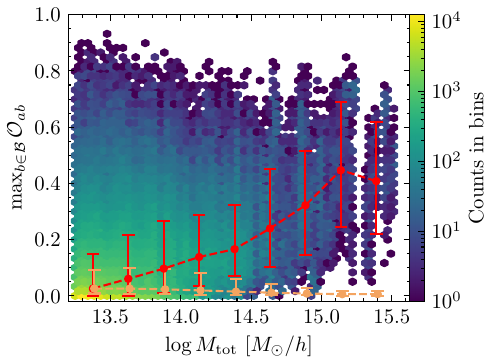}
        \caption{Maximum pair overlaps of a reference halo $\max_{b \in \mathcal{B}} \mathcal{O}_{a b}$ with a single simulation, concatenated across all remaining simulations.}
        \label{fig:max_pair_overlap}
    \end{subfigure}
    \caption{Comparison of Lagrangian patch overlaps in \ac{CSIBORG} and \texttt{Quijote} simulations. In both cases, a single reference simulation is assumed, and for each halo a list of overlaps is calculated per crossing simulation. The $x$-axis denotes the mass of the reference halo, the hex bins illustrate \ac{CSIBORG} overlaps, with the lines delineating overlaps in \ac{CSIBORG} and \texttt{Quijote}, each accompanied by $1\sigma$ error bars showing the spread among points. The differentiation in overlap behaviour between the two simulations becomes evident at high mass, where the constraints of the \ac{CSIBORG} suite become significant. The concatenated pair overlaps show no trend because they are dominated by frequent chance overlaps. On the other hand, there is a clear trend in the maximum overlaps towards higher halo masses.}
    \label{fig:pair_and_pairmax_overlaps}
\end{figure*}

We showcase the framework on the \ac{CSIBORG} suite. We first calculate the overlaps of halo initial Lagrangian patches defined in~\cref{eq:overlap}. We then calculate the halo mass, spin, and concentration of overlapping haloes and compare them to the reference halo. At each step, we compare our findings to the unconstrained \texttt{Quijote} suite to assess the significance of the results. We calculate the overlaps for haloes whose total mass exceeds $10^{13.25}~M_\odot / h$ and pairs of haloes closer in mass than $2~\mathrm{dex}$. The latter is simply to reduce computational expense, since halo pairs with larger differences in mass have negligible overlaps ($\lesssim 1$ per cent).

Although the \ac{DM} particle mass differs between \ac{CSIBORG} and \texttt{Quijote}, this does not prevent us from comparing trends across the two simulation suites. This discrepancy would only be problematic if we were matching a \ac{CSIBORG} box to \texttt{Quijote}, which we do not. The overlap metric itself does not directly depend on the particle mass, as the particles are deposited onto a regular grid: it is primarily influenced by the shot noise resulting from sampling the density field with a finite number of particles. However, we do not expect this to significantly impact our results. A halo with a mass of $10^{13.25}~M_\odot / h$ in \texttt{Quijote} already consists of 200 particles, with the shot noise decreasing further at higher halo masses.


\subsection{Overlaps}\label{sec:overlaps}

We begin by using a single reference simulation and assessing how consistently its haloes are present in the remaining \ac{IC} realisations. We calculate the \emph{non-zero} overlaps between its haloes and those of another simulation following the approach of~\cref{sec:method}. This yields for each halo in the reference simulation a set of overlaps (which could potentially be empty), but its sum will never exceed one because of the denominator of~\cref{eq:intersect}. Retaining the same reference, the process is reiterated across the remaining \ac{IC} realisations.


\subsubsection{Pair overlap}

We present the overlaps between the haloes in~\cref{fig:pair_overlap} where, for every reference halo, we concatenate all sets of overlaps with the remaining \ac{IC} realisations. For comparison, in~\cref{fig:pair_overlap} we also show the overlaps in \texttt{Quijote} where we again use a single reference simulation. The hex-bin shows the \ac{CSIBORG} results and the lines indicate binned medians and $2\sigma$ spread. It is evident that in both \ac{CSIBORG} and \texttt{Quijote} the most likely overlap value is close to zero, which follows from the predominance of non-matched haloes in both simulation suites. However, a notable distinction emerges in \ac{CSIBORG}: there is a pronounced tail of high overlaps, especially evident for haloes above $10^{14}~M_\odot / h$. These massive haloes have counterparts that occupy the same part of the initial snapshot in the majority of the \ac{IC} realisations. This is not the case in \texttt{Quijote}, where in fact the high-overlap tail becomes less significant for more massive haloes due to their rarity.


\subsubsection{Maximum pair overlap}

\begin{figure}
    \includegraphics[width=\columnwidth]{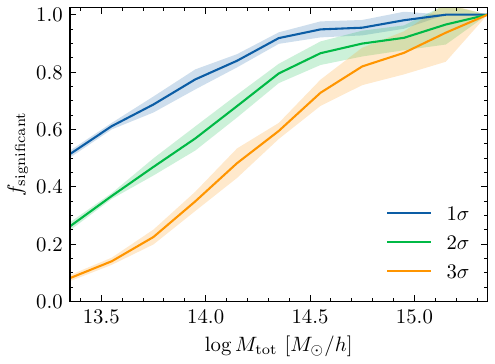}
    \caption{Fraction of \ac{CSIBORG} reference haloes, as a function of $\log M_{\rm tot}$, with a mean maximum overlap with the remaining \ac{IC} realisations exceeding the $1,\,2,\,3\sigma$ thresholds in \texttt{Quijote} ($84.1,\,97.7,\,99.9$ per cent), as shown in~\cref{fig:pair_and_pairmax_overlaps}. The bands show the $1\sigma$ spread among the \ac{CSIBORG} realisations. Even at the lower mass threshold a large proportion of \ac{CSIBORG} haloes has maximum overlaps more significant than in \texttt{Quijote}.}
    \label{fig:maxoverlap_fraction}
\end{figure}

\begin{figure*}
    \begin{subfigure}[t]{0.48\textwidth}
        \includegraphics[width=\linewidth]{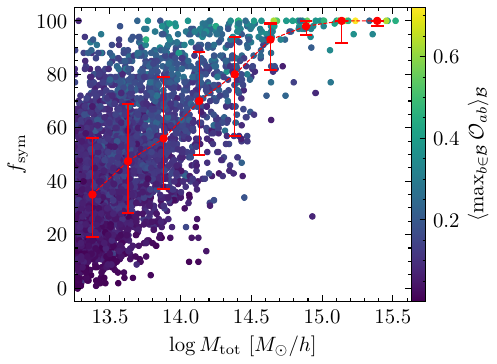}
        \caption{$f_{\rm sym}$ against the reference halo mass.}
        \label{fig:mtot_vs_fsym}
    \end{subfigure}
    \hfill
    \begin{subfigure}[t]{0.48\textwidth}
        \includegraphics[width=\linewidth]{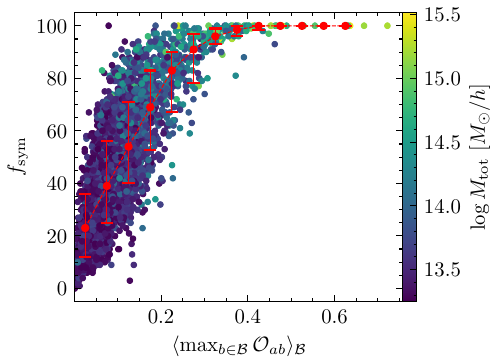}
        \caption{$f_{\rm sym}$ against the average maximum overlap of a halo.}
        \label{fig:maxoverlap_vs_fsym}
    \end{subfigure}
    \caption{Number of simulations in  which a reference halo $a$ and a halo $b$ from a crossing simulation \emph{both} have maximum overlaps with each other, $f_{\rm sym}$, shown for every halo from a single reference simulation in \ac{CSIBORG}. The red line is an arithmetic average in a bin along with $1\sigma$ spread. The highest mass haloes have symmetric maximum overlaps in nearly all \ac{IC} realisations.}
    \label{fig:fsym_plots}
\end{figure*}

Next, we keep the same reference simulation, but instead calculate the maximum overlap of a reference halo in each other \ac{IC} realisation. In~\cref{fig:max_pair_overlap} we present the median of the maximum overlaps for each reference halo across the remaining \ac{IC} realisations. Unlike before, in \ac{CSIBORG} the mean trend of the maximum overlaps as a function of mass is no longer close to zero. On the other hand, in \texttt{Quijote} the mean trend of maximum overlaps remains close to zero. This indicates that there are haloes in any pair of simulations that match well in \ac{CSIBORG}, especially at high mass, while there are not in \texttt{Quijote}. We verify that this conclusion holds irrespective of the choice of reference simulation.

In~\cref{fig:maxoverlap_fraction} we show the fraction of \ac{CSIBORG} haloes in a reference simulation that have a median maximum overlap with other simulations over $1,\,2,\,3\sigma$-level thresholds calculated in \texttt{Quijote} ($84.1,\,97.7,\,99.9$ per cent). This figure again highlights that more massive haloes are more clearly constrained: already at $10^{14} M_\odot / h$ about $75$ per cent of \ac{CSIBORG} haloes have median maximum overlaps more significant than the $1\sigma$ level in \texttt{Quijote}.

In \cref{fig:fsym_plots}, we show in \ac{CSIBORG} for every halo from a single reference simulation the number of simulations in which a reference halo $a$ and a halo $b$ from a crossing simulation \emph{both} have maximum overlaps with each other ($f_{\rm sym}$). We plot $f_{\rm sym}$ against both the reference halo mass and the average maximum overlap of a halo. On average, haloes around $\sim 10^{14} M_\odot / h$ have symmetric maximum overlaps in $50$ per cent of the \ac{IC} realisations. In contrast, haloes around $\sim 10^{15} M_\odot / h$ have symmetric maximum overlaps in nearly all \ac{IC} realisations. On the other hand, the relationship between $f_{\rm sym}$ and the average maximum overlap has less scatter than its relationship with mass. For example, haloes with an average maximum overlap of $\sim 0.2$ have symmetric maximum overlaps in approximately $50$ per cent of the realisations. The results of~\cref{fig:maxoverlap_fraction} and~\cref{fig:fsym_plots} demonstrate that the most massive haloes exhibit reduced variance in \ac{CSIBORG} compared to a suite of unconstrained simulations, highlighting the robustness of our method in identifying consistently constrained haloes across different simulations. However, this does not imply that these haloes are accurately reconstructed relative to real observations; rather, it reflects the internal consistency of the simulation suite.


\subsubsection{Probability of being matched}

\begin{figure*}
    \begin{subfigure}[t]{0.48\textwidth}
        \includegraphics[width=\linewidth]{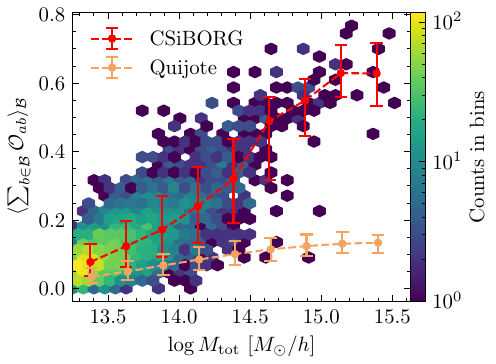}
        \caption{Mean probability of a match averaged over the remaining \ac{IC} realisations.}
        \label{fig:mean_prob_match}
    \end{subfigure}
    \hfill
    \begin{subfigure}[t]{0.48\textwidth}
        \includegraphics[width=\linewidth]{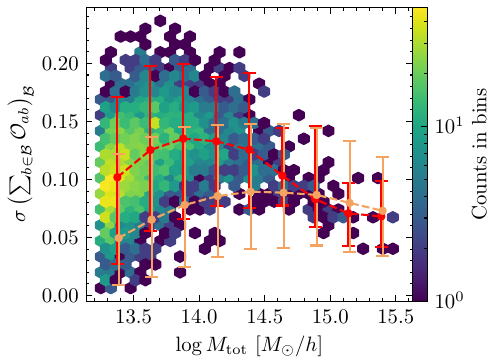}
        \caption{Standard deviation of the probability of a match averaged over the remaining \ac{IC} realisations.}
        \label{fig:std_prob_match}
    \end{subfigure}
    \caption{Probability of a reference halo having a match in the remaining \ac{IC} realisations calculated as the sum of overlaps. The hex bins show the \ac{CSIBORG} data and the lines are the mean trends in \ac{CSIBORG} and \texttt{Quijote}, with $1\sigma$ error bars characterising the spread among points. The probability of a match is typically significantly higher in \ac{CSIBORG} except for the lower mass threshold where the constraints weaken. The truncation at low masses in Fig.~\ref{fig:mean_prob_match}, and the reduction in uncertainty toward very low masses in Fig.~\ref{fig:std_prob_match}, are due to the lower mass threshold of $10^{13.25} M_\odot / h$ when matching haloes.}
    \label{fig:prob_match}
\end{figure*}

Lastly, we calculate the probability that a reference halo has a counterpart in \emph{any} other \ac{IC} realisation, as given by~\cref{eq:prob_match}. In~\cref{fig:prob_match} we plot the mean and standard deviation of this probability, averaged over the remaining \ac{IC} realisations. If a halo has no potential match in another simulation, then the sum of its overlaps in that simulation is simply zero. Mirroring previous results, there is a clear distinction between \ac{CSIBORG} and \texttt{Quijote} above $\sim 10^{14}~M_\odot / h$, in that the majority of \ac{CSIBORG} haloes are matched while the majority of \texttt{Quijote} haloes are not. Below this mass, the distinction weakens, though it remains significant. The uncertainty on the probability of a match shown in~\cref{fig:std_prob_match} peaks at $10^{14}~M_\odot / h$. Below this mass, and particularly near the lower mass threshold of $10^{13.25} M_\odot / h$, there are only haloes with a mass above this threshold to overlap with and not below it, thus underestimating the uncertainty.

This is complementary to the results of~\cref{fig:max_pair_overlap}, which was, instead, sensitive to the maximum overlap a reference halo has with another simulation. On the other hand, in~\cref{fig:prob_match} a high probability of a match can also be due to adding many small overlaps.

In both \ac{CSIBORG} and \texttt{Quijote} there is a trend that more massive haloes have higher sum of overlaps, although this is more pronounced in \ac{CSIBORG}. However, this is not surprising, since the most massive haloes have initial Lagrangian patches of $\sim 10~\mathrm{Mpc} / h$ and thus naturally overlap with more objects. For comparison, see the trend of~\cref{fig:max_pair_overlap} where the mean trend of the maximum overlaps in \texttt{Quijote} never rises---while the large haloes have overlaps with many small ones such that the sum of overlaps may increase, the maximum overlap a given reference halo has with any one halo in a crossing simulation does not increase with mass.

\begin{figure}
    \includegraphics[width=\columnwidth]{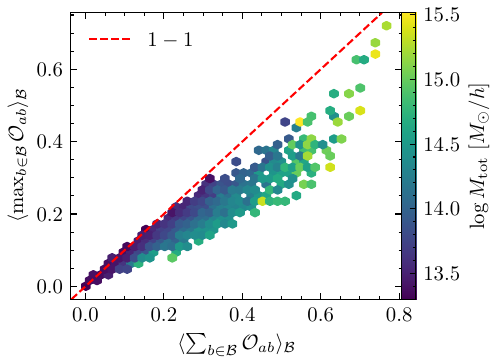}
    \caption{Relation between the mean summed overlap and mean maximum overlap of reference haloes, both of which are averaged over the remaining \ac{CSIBORG} \ac{IC} realisations. Although the most massive haloes deviate from the $1-1$ line due to contributions of smaller overlaps, it is clear that the summed overlaps are typically dominated by the overlap to a single object, making the match relatively unambiguous.}
    \label{fig:summed_to_max_overlap}
\end{figure}

In~\cref{fig:summed_to_max_overlap}, we show in \ac{CSIBORG} the relation between the probability of being matched and the maximum overlap averaged over the \ac{IC} realisations. The two quantities are strongly correlated, though the most massive haloes deviate from the $1-1$ line, as smaller overlaps start to contribute more significantly to the sum over overlaps. Nevertheless, even for the most massive haloes, this shows that sum over overlaps is typically dominated by one object that has a large overlap with the reference halo.

\subsection{Halo properties of overlapping haloes}

Now that we have explored the properties of the overlap statistic and its behaviour across the \ac{CSIBORG} suite, we turn our attention to investigating the properties of haloes that it matches. This includes their separation (\Cref{sec:matched_halo_separation}), mass (\Cref{sec:matched_halo_mass}), peculiar velocity (\Cref{sec:matched_peculiar_velocity}) and concentration and spin (\Cref{sec:matched_halo_spin_concentration}).


\subsubsection{Final snapshot separation of overlapping haloes}\label{sec:matched_halo_separation}

For a reference halo that has a significant overlap with a halo in other \ac{IC} realisation, our anticipation is that their proximity in the \acp{IC} will make the pair unusually close in the final snapshot as well. Because there are nearby haloes between boxes even in the absence of any \ac{IC} constraints, we juxtapose our results with those from the unconstrained \texttt{Quijote} suite. Any suppression in distance observed in \ac{CSIBORG} relative to \texttt{Quijote} can then be ascribed to the constraints.

\begin{figure}
    \includegraphics[width=\columnwidth]{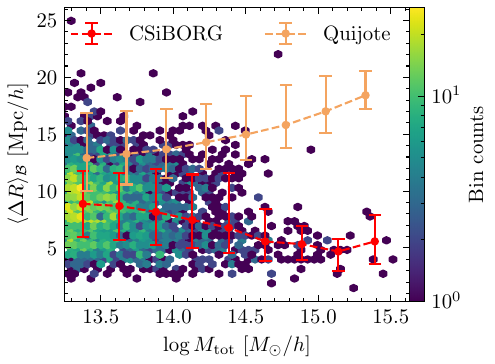}
    \caption{The overlap-weighted mean separation of haloes in the final snapshot $\Delta R$ of \ac{CSIBORG} realisation $7444$ averaged over all remaining \ac{IC} realisations.  The hex bins are the \ac{CSIBORG} overlaps and the lines are the \ac{CSIBORG} and \texttt{Quijote} mean trends, respectively, with $1\sigma$ spread among points Although the halo positions are constrained across the entire mass range when compared to \texttt{Quijote}, there is a significant variation in the positions of the lower mass objects.}
    \label{fig:mass_vs_sep}
\end{figure}

We show the results in~\cref{fig:mass_vs_sep}. We cross-match all \ac{IC} realisations to a single reference simulation and compute $\langle \Delta R\rangle$, the overlap-weighted average separation of haloes in the final snapshot averaged over all \ac{IC} realisations. We calculate the separation as $\Delta R = \lVert(\bm{x}_a - \bm{x}_b)\rVert$, where $\bm{x}_a$ and $\bm{x}_b$ are the position of the reference and matched halo, respectively, and $\lVert \bm{x} \rVert= \sqrt{\bm{x} \cdot \bm{x}}$. In \ac{CSIBORG} haloes are consistently more likely to remain close in the final snapshot if they originate from the same Lagrangian patch. The lowest mass matched objects in \ac{CSIBORG} show a variation of $\sim 10~\mathrm{Mpc} / h$. For an individual reference halo and its maximum-overlap matches from the remaining \ac{IC} realisations we find, as expected, a strong negative correlation between the overlap value and their $z = 0$ separation.

\subsubsection{Mass of matched haloes}\label{sec:matched_halo_mass}

\begin{figure*}
    \includegraphics[width=\textwidth]{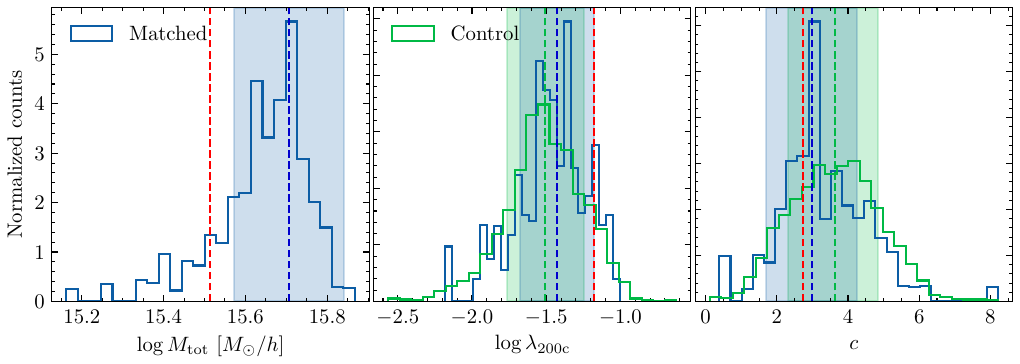}
    \caption{Distributions of most likely matched haloes' properties ($\log M_{\rm tot}$, $\log \lambda_{\rm 200c}$, $c$) with the most massive high-resolution halo in one \ac{IC} realisation (\texttt{7444}) of \ac{CSIBORG} shown as the blue histogram. The green ``control'' histogram is the spin and concentration of haloes with a similar mass from the remaining \ac{IC} realisations: from each we take the $10$ haloes closest in mass to the reference halo. The blue and green vertical lines are the mode of the ``matched'' and ``control'' histograms, respectively, and the corresponding shaded bands are $1\sigma$ uncertainties. The red line is the reference halo property. A comparison to the control distribution reveals that neither the spin nor the concentration is constrained. The matched concentration of this halo has a sharp peak near the reference concentration, however a similar trend is \emph{not} observed for other massive haloes.}
    \label{fig:single_ref_vs_expected}
\end{figure*}

The next quantity that we look at is the most probable mass of the matched haloes. We calculate this following the approach outlined at the end of~\cref{sec:method}. From each crossing simulation, we find the mass of the maximally overlapping halo and construct a weighted histogram, where the weights are the maximum overlaps. We then define the most probable mass as the mode of this distribution, with its uncertainty given by the square root of the overlap-weighted average square of residuals around the mode. We plot an example of this in the left panel of~\cref{fig:single_ref_vs_expected} for the most massive halo in a single \ac{CSIBORG} realisation, finding the most likely mass to be within $0.2~\mathrm{dex}$ of the reference halo mass. This is in part by construction, since the overlap itself is preferentially higher for haloes that have a similar mass. However, while the overlap is preferentially higher for haloes of a similar mass, it does not guarantee the objects being a ``good'' match. If that were the case, then we would find that even in \texttt{Quijote} the mass of the matched haloes is on average close to the mass of the reference halo, which it is not.

\begin{figure*}
    \includegraphics[width=\textwidth]{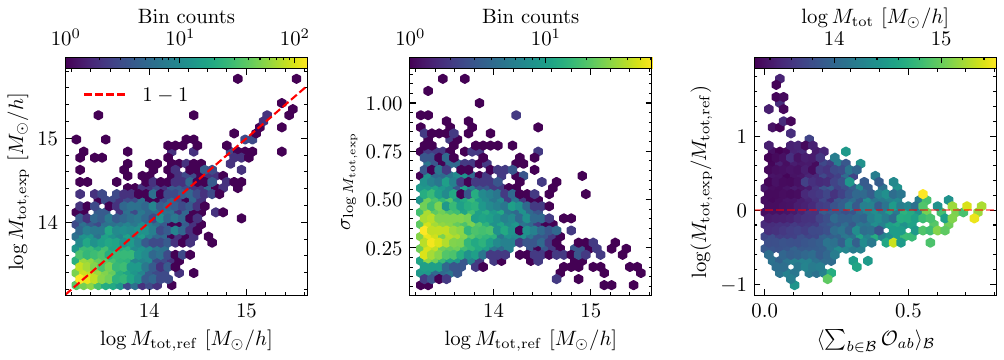}
    \caption{Most likely mass of haloes matched to a single realisation of \ac{CSIBORG} averaged over all crossing simulations and defined as the mode of a histogram such as~\cref{fig:single_ref_vs_expected}. The \emph{left} panel shows the reference-to-expected halo mass relation, the \emph{middle} panel shows the uncertainty of the expected mass and the \emph{right} panel shows the ratio of the expected mass to the reference mass as a function of probability of being matched defined in~\cref{eq:prob_match}. The agreement between the reference and matched mass is strongly correlated with the matching probability.}
    \label{fig:reference_vs_expected_mass}
\end{figure*}

By analysing histograms like~\cref{fig:single_ref_vs_expected}, we calculate and show in~\cref{fig:reference_vs_expected_mass} the most likely mass of all reference haloes in \ac{CSIBORG} above $10^{13.25}~M_\odot / h$. In this figure, we show three panels: comparison between the reference and most likely halo mass, the uncertainty of the most likely mass as a function of the reference mass and the ratio of the most likely mass to the reference mass as a function of the median probability of being matched.

The reference and most likely matched halo masses agree well for the most massive objects, with deviations from the $1-1$ line increasing towards lower mass. However, these objects also have a consistently lower probability of being matched (right panel of~\cref{fig:reference_vs_expected_mass}) and, therefore, it is not unexpected. Below the scale of $\sim 10^{14}~M_\odot / h$ the matching is less robust: we only cross-match haloes with mass above $10^{13.25}~M_\odot / h$ since below this scale the \ac{IC} constraints weaken as indicated by~\cref{fig:max_pair_overlap} and~\cref{fig:mean_prob_match}. Therefore, a reference halo near the threshold will only have potential matches that are above this threshold, thus biasing the matching procedure.

\subsubsection{Peculiar velocity alignment}\label{sec:matched_peculiar_velocity}

\begin{figure}
    \includegraphics[width=\columnwidth]{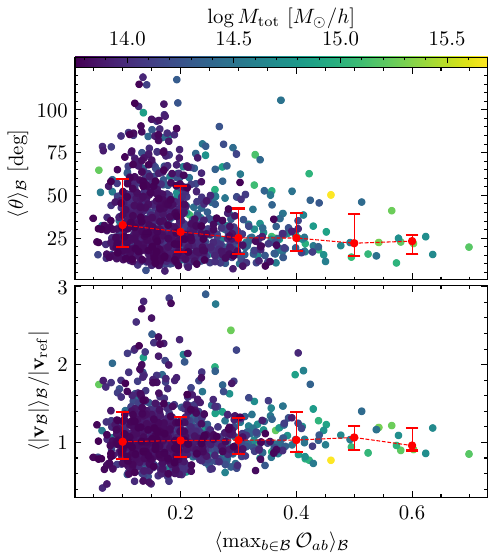}
    \caption{The mean alignment angle (\emph{top} panel) and ratio of magnitudes (\emph{bottom} panel) of reference haloes with the remaining \ac{IC} realisations in \ac{CSIBORG}, plotted as a function of the mean maximum overlap of the reference halo averaged over the \ac{IC} realisations. The red line is the mean trend in a bin with $1\sigma$ spread among points. Higher overlapping haloes tend to have their peculiar velocities more aligned in final snapshot.}
    \label{fig:velocity_alignment}
\end{figure}

In~\cref{fig:velocity_alignment}, we show the alignment and ratio of the magnitudes of the peculiar velocities in the final snapshot of \ac{CSIBORG}. We define the alignment angle $\theta$ between the peculiar velocity vectors of haloes $a$ and $b$ with peculiar velocities $\bm{v}_a$ and $\bm{v}_b$, respectively, as:
\begin{equation}
    \cos \theta = \frac{\bm{v}_a \cdot \bm{v}_b}{|\bm{v}_a|~|\bm{v}_b|}.
\end{equation}

For each reference halo we calculate the alignment of its peculiar velocity vector with the corresponding vector of the highest overlapping halo from another \ac{IC} realisation. We find that the alignment of a reference halo and a single maximum overlap crossing halo is strongly correlated with the magnitude of this overlap across the \ac{IC} realisations. The alignment and ratio of the magnitudes plotted in~\cref{fig:velocity_alignment} is averaged over the crossing \ac{IC} realisations. While most matched cluster-mass haloes show strong alignment in the final snapshot, there are exceptions. These often coincide with a lower mean maximum overlap. Nevertheless, even plotting the alignment as a function of the mean maximum overlap, there remains a significant scatter. We also compare the magnitudes of peculiar velocities, finding their ratios to be on average $\sim 1$, however, with significantly larger scatter towards smaller overlaps.

\subsubsection{Spin and concentration constraint}\label{sec:matched_halo_spin_concentration}

Lastly, we turn our attention to the spin and concentration of these haloes. We use the Bullock spin definition
\begin{equation}\label{eq:bullock_spin}
    \lambda_{\rm 200c} = \frac{J_{\rm 200c}}{\sqrt{2} M_{\rm 200c} V_{\rm 200c} R_{\rm 200c}},
\end{equation}
where $J_{\rm 200c}$ is the angular momentum magnitude of particles within $R_{\rm 200c}$ and $V_{\rm 200c}^2 = G M_{\rm 200c} / R_{\rm 200c}$~\citep{Bullock_2001}. We define concentration as the ratio of the virial radius to the scale radius of the Navarro-Frenk-White (NFW) profile, $c = R_{\rm 200c} / R_{\rm s}$~\citep{Navarro_1996}.

We calculate the most likely spin and concentration of the matched haloes following the approach outlined in~\cref{sec:method}. In the middle panel of~\cref{fig:single_ref_vs_expected} we show the comparison of the spins of overlapping haloes to the spin of a reference halo, which we take to be the most massive halo in one \ac{IC} realisations of \ac{CSIBORG}. However, unlike previously, we do not find any preference for the spin of overlapping haloes to be similar to the reference halo spin, regardless of whether we weight the matched haloes by overlap or not. The matched distribution in~\cref{fig:single_ref_vs_expected} has a secondary peak near the reference spin; however, it is not statistically significant. In fact, the distribution of the matched spins is in good agreement with the simulation average, which is approximately a mass-independent Gaussian distribution in $\log \lambda_{\rm 200c}$. We find similar conclusions to hold for all haloes, regardless of their mass.

Next, we investigate the concentration of the matched haloes. In the right panel of~\cref{fig:single_ref_vs_expected} we show the comparison of the concentrations of overlapping haloes to the concentration of a reference halo, which we again take to be the most massive halo in one \ac{IC} realisations of \ac{CSIBORG}. We find that in this particular example, the mode of the weighted distribution agrees well with the reference halo concentration, but the width of this distribution is still similar to the expectation from the simulated mass--concentration relation. To delve deeper, we assess the matched concentrations for all haloes with a mass exceeding $10^{15}~M_\odot / h$, which we previously identified as being consistently reconstructed. We find no discernible correlation between the most likely and reference concentrations.  We also find that the concentration of the highest overlap halo from each \ac{IC} realisation is within $10$ ($30$) per cent of the reference value in only $20$ ($50$) per cent of realisations, without any clear dependence on the reference halo mass. The agreement seen in~\cref{fig:single_ref_vs_expected} is thus coincidental or specific to the halo in question. However, even in such instances, the significance is questionable as there is no notable improvement over the mass--concentration relation.


\section{Discussion}\label{sec:discussion}


\subsection{Interpretation of the results}

Constrained simulations enable an object-by-object comparison of the local Universe with theory. If a direct, or at least probabilistic, correspondence between a simulated halo and an observed structure can be established, constrained simulations potentially allow inference of the properties and assembly histories of nearby objects and hence pose rigorous tests for galaxy formation models and cosmology. In this work, we outline and test a method for assessing whether a halo is robustly reconstructed across a set of simulations, differing for example in \acp{IC} drawn from the posterior of a preceding inference. This allows us to compare the properties of the ``same'' haloes across the simulations. By ``robustly reconstructed,'' we refer specifically to the suppression of variance in halo properties across the sampled \acp{IC}, meaning the haloes originate from the same Lagrangian patches and have similar properties at $z = 0$. An important additional aspect in assessing the quality of any local Universe reconstruction is how faithfully it reproduces the observed Universe. However, we do not test this aspect in the present work.

We illustrate our framework by applying it to the \ac{CSIBORG} suite of $101$ constrained simulations of the local $\sim 155~\mathrm{Mpc} / h$ Universe with \acp{IC} on a grid of spacing $2.65~\mathrm{Mpc} / h$ derived from the \ac{BORG} inference of the $\mathrm{2M}\texttt{++}$ catalogue. We find that cluster-mass haloes (\(M \gtrsim 10^{14}~M_\odot / h\)) are consistently reconstructed across the suite and originate from the same Lagrangian region. Haloes of this mass are distributed over $\sim70$ cells in the initial snapshot at $z = 69$. Assuming that the comoving cell density in the initial snapshot is $\Omega_{\rm m} \rho_{\rm c}$, where $\rho_{\rm c}$ is the current critical density of the Universe, we can approximate the spatial resolution, $L$, required to distribute a mass $M$ over $N$ cells as:
\begin{equation}
    L \approx 2.6~\mathrm{Mpc} / h \left(\frac{M}{10^{14}~M_\odot / h} \frac{70}{N}\right)^{1/3}.
\end{equation}
Assuming the criterion of $70$ resolution elements across the initial Lagrangian patch to be universal, this suggests that for haloes of mass $10^{15},\,10^{14},\,10^{13}~M_\odot / h$ to be robustly reconstructed one must use \ac{IC} constraints at the scales $5.5,\,2.6$ and $1.2~\mathrm{Mpc} / h$, respectively.

While high-mass haloes are consistently reconstructed and have counterparts of similar mass and peculiar velocity across all \ac{IC} realisations, the overlapping haloes originating from same Lagrangian regions are not similar in neither their spin or concentration. Despite the \ac{BORG} reconstruction employed in this work not utilising a peculiar velocity catalogue as a constraint---relying solely on galaxy positions in the $\mathrm{2M}\texttt{++}$ catalogue---it is reassuring to find that the highest mass haloes indeed have aligned velocities, since to some extent galaxy positions are complementary to the peculiar velocity field information. In fact,~\ac{BORG} has been used in the past to reconstruct the local peculiar velocity field~\citep{Jasche2019_BORG}.

\cite{Cadiou_2021} demonstrated that the angular momentum of haloes can be accurately predicted directly from their Lagrangian patches, but that it exhibits chaotic behaviour under small changes to the patch boundary. Although the high-mass \ac{CSIBORG} haloes are present in all \ac{IC} realisations, their overlaps never reach unity and so their Lagrangian patches are not exactly aligned. Therefore, given the findings of~\citeauthor{Cadiou_2021}, the lack of constraint on spin is not surprising. The halo concentration depends on both the Lagrangian patch configuration and subsequent accretion history~\citep{Rey_2019}. While we find that on average the halo concentration in~\ac{CSIBORG} is not constrained, we leave a detailed examination of specific observed clusters for future work along with comparing the mass accretion histories of haloes that are initially strongly overlapping. This will help tease out the properties of haloes and their constituent particles that are responsible for setting their concentrations.

For certain haloes, we observe that those originating from consecutive \ac{CSIBORG} simulations tend to have higher maximum overlaps. \ac{CSIBORG} resimulates every $24$\textsuperscript{th} step of the \ac{BORG} chain, yet the autocorrelation length of the chain is $\sim 100$. \ac{CSIBORG}  thus effectively over-samples the \ac{BORG} posterior---even though it varies without correlation the unconstrained small-scale modes at each step---which could lead to an overestimation of confidence in the \ac{BORG} constraints if relying only on a few consecutive \ac{CSIBORG} samples. To mitigate this effect, we have averaged across all remaining $100$ \ac{IC} realisations for each reference realisation, the majority of which are fully decorrelated from it. A fully satisfactory solution would be to resimulate only decorrelated \ac{IC} realisations, however it is challenging to derive a large number of these due to the high computational cost of the \ac{BORG} inference.


\subsection{Comparison with the literature}\label{sec:literature_comparison}

\cite{Max2022} introduce an algorithm to assess whether ``twins'' of a single halo can be identified in all \ac{IC} realisations of a constrained simulation. This is done by selecting a reference simulation and cataloguing the positions of all haloes. A halo is then chosen from the reference simulation, and its size is calculated as $R_{\rm 200c} = \left[ (3 M_{\rm 200c}) / (4\pi \cdot 200 \cdot \rho_{\rm crit}) \right]^{1/3}$. In the other simulations, the haloes within a designated ``search radius'' of the reference halo are then identified and the one most similar in mass selected. If no haloes are found within the search radius, the reference halo is discarded as not present within the crossing simulation. This approach differs from ours in several important ways:
\begin{enumerate}
    \item we match haloes directly in the \acp{IC}, which are constrained by \ac{BORG}, instead of relying on the forward-modelled realisations of the final snapshot,
    \item we do not require the matched haloes to be within a fixed radius of one another, but instead calculate the overlap of their initial Lagrangian patches,
    \item while the procedure of \citeauthor{Max2022} is inherently binary, ours has a continuous interpretation in terms of probabilities.
\end{enumerate}

\begin{figure}
    \includegraphics[width=\columnwidth]{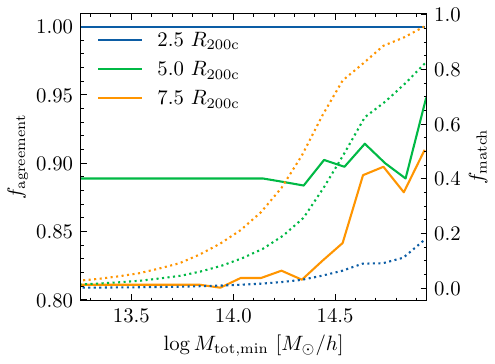}
    \caption{Comparison of our matching procedure to that of~\protect\cite{Max2022}, who identify ``twin'' haloes in the final snapshots of \ac{CSIBORG}. Solid lines correspond to $f_{\rm agreement}$, which is the fractional agreement between the~\citeauthor{Max2022} matches and the maximum-overlap pairs above mass $M_{\rm tot, min}$. Dotted lines are $f_{\rm match}$, the fraction of haloes above $M_{\rm tot, min}$ identified as matches by~\citeauthor{Max2022}. The high $f_{\rm agreement}$ indicates excellent agreement between the approaches regardless of the search radius (shown in the legend). However, only a small fraction of haloes are identified as matches by~\citeauthor{Max2022}. Our method establishes a match probability applicable even to far more uncertain matches.}
    \label{fig:comparison_to_max}
\end{figure}

It is nevertheless instructive to compare our results. In~\cref{fig:comparison_to_max} we show the fraction of matches identified via the approach of~\citeauthor{Max2022} that are also the maximum-overlap pairs for several choices of search radii in \ac{CSIBORG}. Due to the stringent requirement of~\citeauthor{Max2022} that a halo must have a ``twin'' in \emph{all} \ac{IC} realisations, there is a strong agreement between the two approaches. We calculate $f_{\rm agreement}$, the fraction of matches identified by~\citeauthor{Max2022} that are also the maximum overlap pairs as a function of a lower mass threshold. For the fiducial search radius used by~\citeauthor{Max2022} the agreement is $\sim 90$ per cent regardless of the mass threshold. However, this comes at the cost of only a small number of haloes being identified as matches by~\citeauthor{Max2022}: for example, at $10^{14}~M_\odot / h$ this is only $\sim 10$ per cent of the haloes. Our method has the advantage of providing useful information even in the regime where the matching across all realisations is partial or even weak.

Another methodology with similarities to ours is that of~\cite{LUM_2023}, in which the goal is to identify the simulation most representative of the local Universe. However, a crucial difference between our work and that of \citeauthor{LUM_2023} is that we focus on studying the consistency of constraining haloes across the posterior samples of an \ac{IC} inference, rather than assessing the reliability of matching these haloes to the local Universe. They use a Wiener filter-based reconstruction that must be supplied with random small-scale/unconstrained modes. This strategy can alternatively be conceptualised as an intensive fine-tuning process to pinpoint the optimal random seed on a very fine grid level, which can then be re-simulated~\citep{McAlpine_2022_Sibellius}. \citeauthor{LUM_2023} find the realisation that best resembles the local Universe by minimising the sum of $p$-values of simulated cluster-mass haloes being at the locations of observed clusters. However, while they find the most representative constrained simulations, their approach does not provide any information about the consistency with which the matched simulated haloes are reconstructed. We believe that this is crucial information to determine the confidence with which we can assert the properties of the dark matter distribution of the local Universe, given the quality and quantity of data used to set the constraints.

A similar technique to match observed galaxies to haloes from constrained cosmological simulations has been applied by~\cite{Zhang_2022,Xu_2023}. \citeauthor{Zhang_2022} develop a ``neighbourhood'' \ac{SHAM}, a \ac{SHAM} model specifically tailored for constrained cosmological simulations which ranks haloes based on both their peak mass and closeness in position and velocity to the observed galaxy~\citep{Yang_2018}. This statistical approach enables them to connect haloes in their single constrained simulation with observed galaxies, thereby facilitating the study of, e.g., the galaxy-to-halo size relation. Our method would allow folding in the uncertainties associated with the reconstruction.


\section{Conclusion}\label{sec:conclusion}

We have investigated the extent to which haloes are robustly reconstructed across multiple cosmological simulations that sample the \ac{IC} posterior of a previous inference. While this question is particularly pertinent to simulation suites constrained (with uncertainty) to match the local Universe, other applications include the matching of haloes between \ac{DM}-only simulations and their hydrodynamical counterparts, or simulations of varying cosmology such as $\Lambda\mathrm{CDM}$ vs modified gravity.

We argue that for a halo to be consistently reconstructed, its Lagrangian patch in the initial conditions must strongly overlap with a halo in most other realisations of the suite, a condition implying similarity in both mass and location. This is a stricter and more causal measure than similarity in the final snapshot, providing a clearer condition for two haloes to be ``the same''. In the future, an even stricter measure of similarity may be introduced by measuring the haloes' initial overlap in the $6\mathrm{D}$ phase space directly, instead of only the $\mathrm{3}\mathrm{D}$ position space.

We apply the method to \ac{CSIBORG}, a suite of constrained simulations with initial conditions sampled from the posterior of the \ac{BORG} algorithm applied to the $\mathrm{2M}\texttt{++}$ galaxy number density field. We establish the significance of the results by comparing to the unconstrained \texttt{Quijote} suite. Based on the criteria mentioned above, we find that cluster-mass haloes ($M \gtrsim 10^{14} M_\odot / h$) are consistently reconstructed in~\ac{CSIBORG} in position, mass, and peculiar velocity, with higher mass haloes typically more strongly constrained. For haloes below this mass threshold, the constraints diminish, and haloes do not consistently originate from the same Lagrangian patches. Regarding secondary halo properties such as spin and concentration, even consistently matched, high-mass haloes display variations across the \ac{IC} realisations. The absence of constraints on concentration beyond the mass-concentration relation is surprising given its strong dependence on mass assembly history. This might imply that the assembly history of even the most massive haloes in \ac{CSIBORG} remains unconstrained; however, we defer a comprehensive study of the mass assembly history to future work. In the future, we will also investigate the extent to which halo overlap correlates with reliability of match to an observed object in the local Universe. In sum, our framework provides a step towards the goal of identifying the \acp{IC} that led to the observed Universe and the objects it contains.

\section{Data availability}

The code underlying this article is available at \url{https://github.com/Richard-Sti/csiborgtools_public} and other data will be made available on reasonable request to the authors.

\section*{Acknowledgements}

We thank Jens Jasche, Guilhem Lavaux, Francisco Villaescusa-Navarro and Tariq Yasin for useful inputs and discussions. We also thank Jonathan Patterson for smoothly running the Glamdring Cluster hosted by the University of Oxford, where the data processing was performed. This work was done within the Aquila Consortium.\footnote{\url{https://aquila-consortium.org}}

RS acknowledges financial support from STFC Grant No. ST/X508664/1, HD is supported by a Royal Society University Research Fellowship (grant no. 211046). This project has received funding from the European Research Council (ERC) under the European Union's Horizon 2020 research and innovation programme (grant agreement No 693024). This work was performed using the DiRAC Data Intensive service at Leicester, operated by the University of Leicester IT Services, which forms part of the STFC DiRAC HPC Facility (www.dirac.ac.uk). The equipment was funded by BEIS capital funding via STFC capital grants ST/K000373/1 and ST/R002363/1 and STFC DiRAC Operations grant ST/R001014/1. DiRAC is part of the National e-Infrastructure. For the purpose of open access, we have applied a Creative Commons Attribution (CC BY) licence to any Author Accepted Manuscript version arising.

\bibliographystyle{mnras}
\bibliography{ref}


\bsp
\label{lastpage}
\end{document}